\documentclass[rnote,a4paper]{aa83}
\usepackage[varg]{txfonts}
\usepackage{graphicx}
\usepackage{psfrag}
\usepackage{gensymb}

\newcommand{\Planck}{\textsc{Planck}}
\newcommand{\boom}{BOOMERanG}

\usepackage[version-1-compatibility]{siunitx} % esempio: $\SI{140}{\giga\hertz}$

\begin{document}

\bibpunct{(}{)}{;}{a}{}{,} % to follow the A&A style

\title{Optimal cosmic microwave background map-making in the presence of cross-correlated noise} 
\subtitle{}
\author{G. de~Gasperis \inst{\ref{inst1},\ref{inst2}}
\and A. Buzzelli \inst{\ref{inst2},\ref{inst3}}
\and P. Cabella \inst{\ref{inst1},\ref{inst2}}
\and P. de~Bernardis \inst{\ref{inst3},\ref{inst4}}
\and N. Vittorio \inst{\ref{inst1},\ref{inst2}}}
\institute{Dipartimento di Fisica, Universit\`a di Roma ``Tor~Vergata'', via della Ricerca Scientifica 1, I-00133, Roma, Italy \label{inst1}
  \and Sezione INFN Roma~2, via della Ricerca Scientifica 1, I-00133, Roma, Italy \label{inst2}
  \and Dipartimento di Fisica, Sapienza Universit\`a di Roma, piazzale Aldo Moro 5, I-00185, Roma, Italy \label{inst3}
  \and Sezione INFN Roma~1, piazzale Aldo Moro 5, I-00185, Roma, Italy \label{inst4}}
\offprints{giancarlo.degasperis@roma2.infn.it}
\date{Received / Accepted}

\abstract
{}
{We present an extension of the ROMA  map-making algorithm for the generation of optimal cosmic microwave background polarization maps. 
The new code allows for a possible cross-correlated noise component among the detectors of a CMB experiment. A promising application is the forthcoming LSPE balloon-borne experiment, which is devoted to the accurate observation of CMB polarization at large angular scales.}
{We generalized the noise covariance matrix in time domain to account for all the off-diagonal terms due to the detector cross-talk.
Hence, we performed preliminary forecasts of the LSPE-SWIPE instrument.}
{We found that considering the noise cross-correlation among the detectors results in a more realistic estimate of the angular power spectra. In particular, the extended ROMA algorithm 
has provided a considerable reduction of the spectra error bars. We expect that this improvement could be crucial in constraining the B-mode polarization at the largest scales.} 
{}

\authorrunning{G. de~Gasperis et~al.}

\titlerunning{Optimal CMB map-making in the presence of cross-correlated noise}


\maketitle

\section{Introduction}\label{intro}

The temperature and polarization patterns of cosmic microwave background (CMB) have been an invaluable source of cosmological information.
 While polarization E-modes have been widely observed and analyzed (e.g., Planck:~\citealt{2015arXiv150702704P}), B-modes are still buried into foreground signal and experimental noise \citep[e.g.,][]{2015PhRvL.114j1301B}. Nowadays, the hunting for B-modes represents one of the most relevant and exciting research fields in cosmology.
In fact, a B-mode detection in CMB polarization would provide a definitive confirmation of the existence of a gravitational waves primordial background, as expected in the inflation paradigm \citep{1999PhR...314....1L}.

Because of the low signal-to-noise ratio, polarization observations require large multidetector arrays. 
The increased detector array size and the integration of many pixels on the same wafer rise the problem of cross talk among the detectors. This can be due to either focal plane temperature variations and/or atmospheric fluctuations. Both these effects are expected to produce common-mode noise in the detectors.
This noise cross-correlation has to be properly taken into account in the map-making procedure.

After the pioneering work of \citet[][]{1996astro.ph.12006W},
map-making has been thoroughly studied in the literature \citep[see e.g.,][and references therein]{2010ApJS..187..212C}.
However, cross-correlated noise among the detectors has not been discussed as much and is crudely neglected. The only detailed treatment that allows us to take  common-mode noise into account explicitly and properly is that by \citet[][]{2008ApJ...681..708P},
for intensity measurements only. Given the pressing interest on primordial B-mode detection, a specific treatment of noise cross-correlation in polarization measurements is now necessary.

Among the next generation of CMB experiments, we focus on the 
%\pdfmarkupcomment[markup=Highlight,color=yellow, author=GdG]{ 
%\textbf
%{
Large-Scale Polarization Explorer\footnote{http://planck.roma1.infn.it/lspe}
%}
(LSPE)
%}{Here we kept capitalization of the experiment official name. We added a footnote with a link to the LSPE website.}
balloon mission, which is devoted to accurate observations of CMB polarization at large angular scales \citep[see e.g.,][]{2012SPIE.8452E..3FD}. The LSPE is expected to improve the limit on the ratio of tensor-to-scalar perturbation amplitudes down to $r \lesssim 0.03$.

In this work we present an extension of the ROMA map-making code \citep[Roma optimal map-making algorithm,][]{2005A&A...436.1159D}
to produce optimal CMB polarization maps out of cross-correlated multi-detector CMB observations.
In particular, we discuss how a proper inclusion of the detector noise cross-correlations results in more realistic estimates of the angular power spectra error bars. This benefit could be crucial for the detection of a primordial B-mode signal at low multipoles.

\section{Formalism, algebra and noise model}\label{algebra}
A CMB experiment observes the sky at a given resolution (i.e. with $N_{p}$ pixels in the sky) and collects $N_{d}$ samples in a given temporal sequence, the time ordered data (TOD).
Once the data are calibrated, any artifact is flagged out and the pointing is reconstructed, the first step in the CMB data analysis is the estimation of optimal sky maps from the TOD.
In the literature two main strategies are present: a maximum-likelihood (minimum-variance) approach \citep[see e.g.,][and references therein]{2005A&A...436.1159D} and destriping techniques \citep[see e.g.,][and references therein]{2011A&A...534A..88T}.

We take the cross-correlated noise component among detectors into account in the following way. 
 The observational data from one detector can be modeled as
\begin{equation}
\textbf{D} = \textbf{A}\textbf{S} + \textbf{n}. \label{D}
\end{equation}
Here, $\textbf{D}$ is the TOD, $\textbf{A}$ is the generalized pointing matrix, 
$\textbf{S}=(I,Q,U)$ is the map triplet and $\textbf{n}$ is the instrumental noise, which accounts for any systematic effects, cosmic ray hits, etc. When dealing with multiple detector observations, the TODs from individual detectors are simply concatenated end-to-end.
Under the assumption of a Gaussian and stationary noise, the generalized least squared approach yields the following maximum-likelihood estimator for the signal
\begin{equation}
 \mathbf{\widetilde S} = \left( \mathbf{A}^T
\mathbf{N}^{-1} \mathbf{A}\right)^{-1}
  \mathbf{A}^T \mathbf{N}^{-1} \mathbf{D}.
\end{equation}\label{GLS}
The noise covariance matrix in the time domain, $\textbf{N} \equiv \left\langle \textbf{n}_t \textbf{n}_{t'} \right\rangle$, is block diagonal only in the case of no cross-correlation among samples of different detectors. The present version of the ROMA algorithm takes into account all the off-diagonal terms, which have usually been neglected in past works. The solution of Eq. \ref{GLS} is a very computationally demanding task because of the large size of the matrix $\textbf{N}$, $kN_p\times kN_p$, where $k$ is the number of detectors. Our algorithm adopts a Fourier-based, preconditioned conjugate-gradient iterative method.

The choice of the noise model is clearly a crucial issue. In general, the detector noise spectrum is in the sum of a stationary Gaussian white noise at higher frequencies and a $1/f$ component at lower frequencies,  which implies an unavoidable correlation among different samples in the single detector time stream. A further complication is the presence of the noise common-mode seen by all the detectors. Its contribution has been estimated directly from data \citep[see e.g.,][]{2006A&A...458..687M}. 
The cross-correlated noise is typical of ground- and balloon-based experiment, as it is sourced mainly by atmospheric fluctuations. However, it could also affect space missions, due to common-mode detector temperature drifts. This work is addressed to the balloon regime, but our treatment is completely general.

%%%%%%%%%%%%%%%%%%%%%%%%%%%%%%%%%%%%%%%%%%%%%%%%%%%%%%%%%%%%%%%%%%%%%%%%%%%%%%%

\section{Forecasts of the LSPE-SWIPE experiment}\label{LSPE}
In this work we focus on the forthcoming LSPE balloon-borne experiment that is devoted to the accurate measurement of CMB polarization at large angular scales.

We use the level of noise cross-correlation among the CMB polarimeters estimated for BOOMERanG \citep[][]{2006A&A...458..687M} as a benchmark for our simulations.
Since the deep similarities between BOOMERanG and the LSPE-SWIPE experiment (long duration stratospheric bolometric experiments subject to atmospheric common-mode fluctuations), we found it natural to use the BOOMERanG noise properties for the LSPE-SWIPE preliminary forecasts.

\subsection{The instrument}

During its circumpolar Arctic flight, LSPE will scan the sky by spinning around the local vertical, while keeping the telescope elevation constant for long periods, in the range 30 to 55 deg. The azimuth scan speed will be set around 2 rpm, i.e. 12 deg/s. A large portion of the northern sky (around 25\% of the celestial sphere) is expected to be observed with an angular resolution of about 1.5 degrees FWHM.
\begin{figure}
\centering
\includegraphics[angle=90,width=\columnwidth]{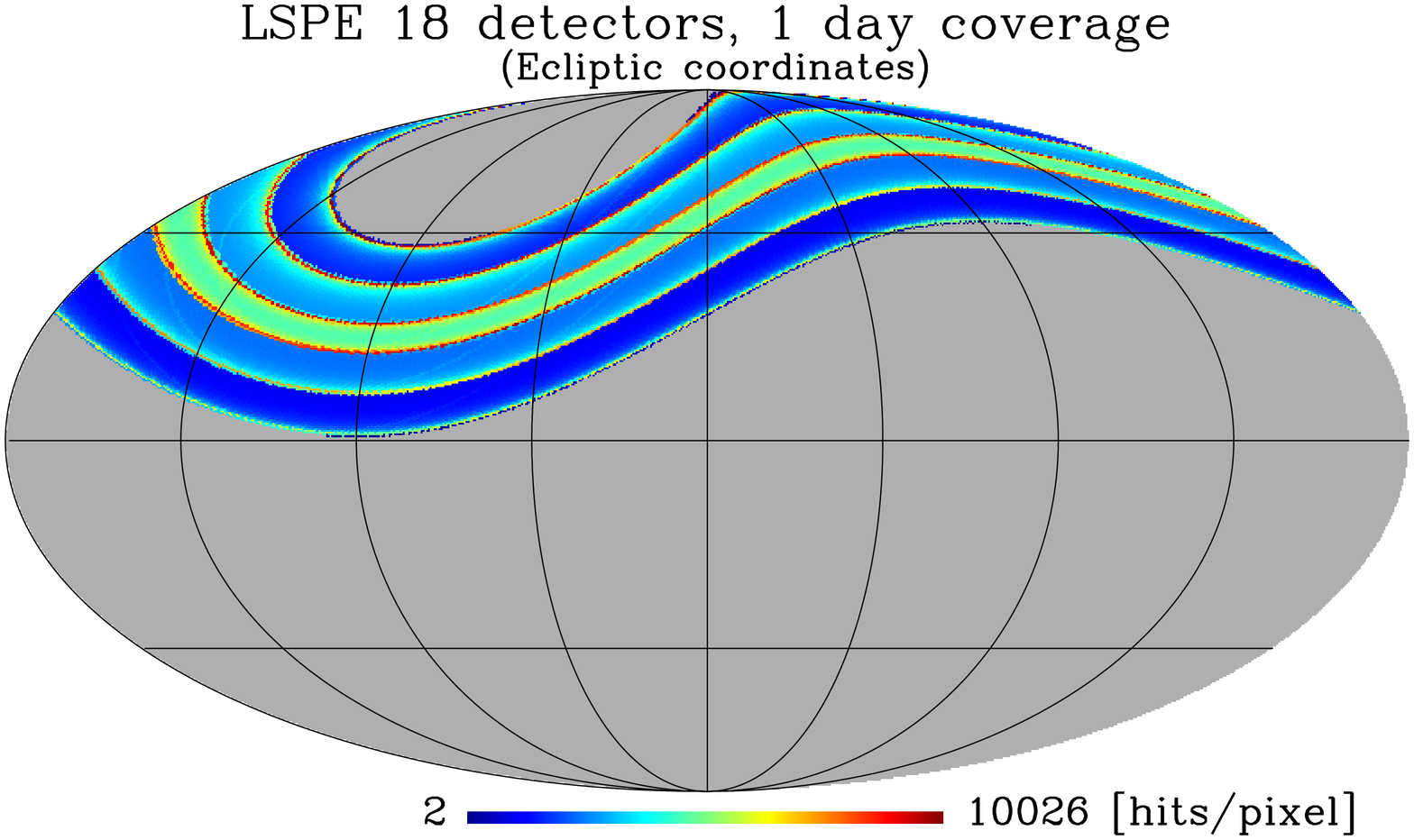}
\includegraphics[angle=90,width=\columnwidth]{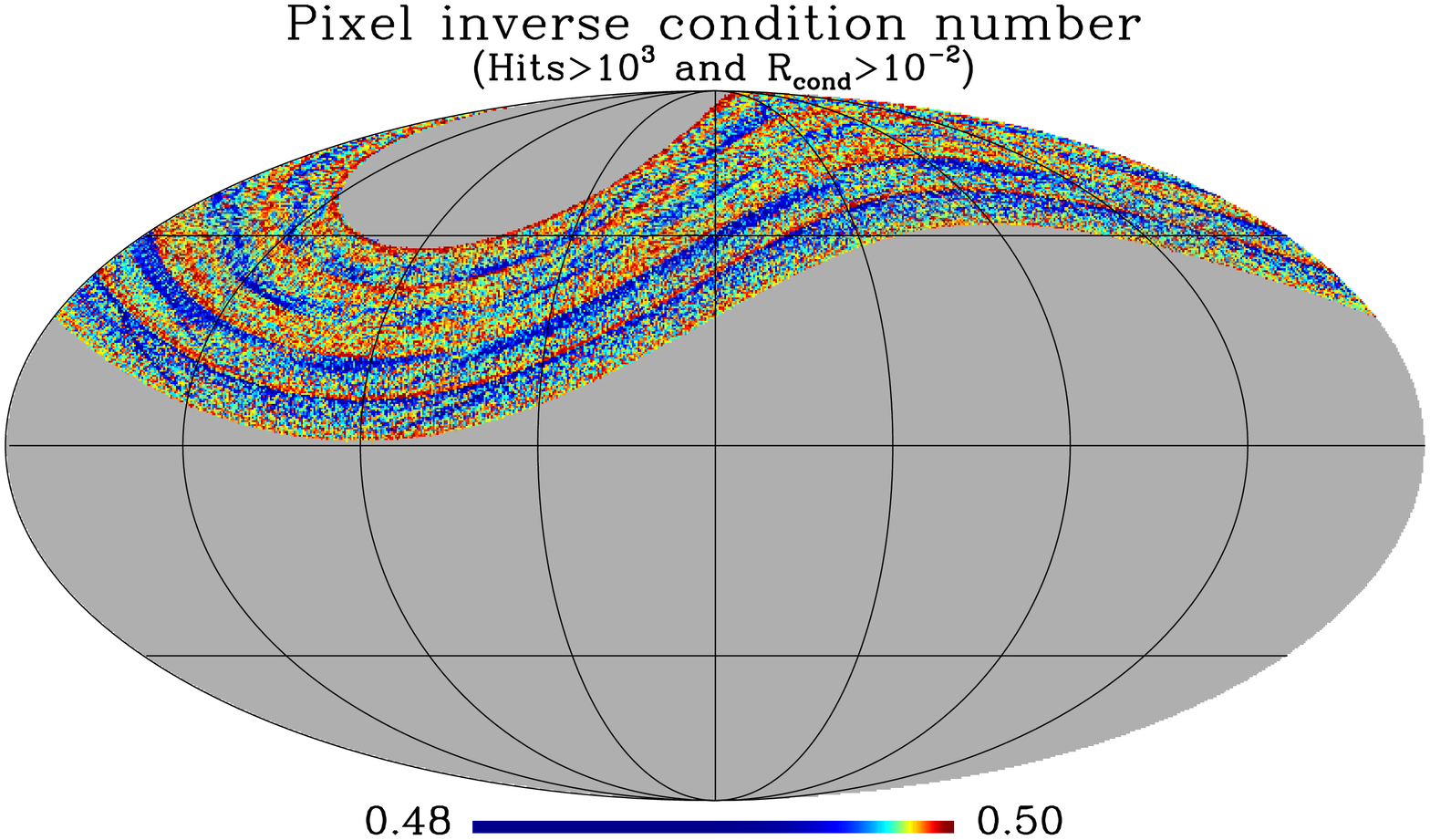}
\caption{Sky region scanned by 18 sample detectors, sparsely located in the two focal planes of LSPE-SWIPE,
in ecliptic coordinates (upper panel)
and the inverse pixel condition number $R_{cond}$ as an estimator of polarization angle coverage per pixel
(lower panel). In our definition, $R_{cond}=1/2$ in case of perfect angle coverage uniformity.
Both maps are at HEALPix~\citep[see][]{2005ApJ...622..759G} $\mathrm{N_{side}}=128$.}
\label{LSPEmask}
\end{figure}
\begin{figure}
\centering
\psfrag{Freq/[Hz]}[c][][3]{Frequency $/[\si{\hertz}]$}
\psfrag{Power/[uK^2/Hz]}[c][][3]{Power $/[\si{\micro\kelvin\squared \per\hertz}]$}
\includegraphics[angle=90,width=\columnwidth]{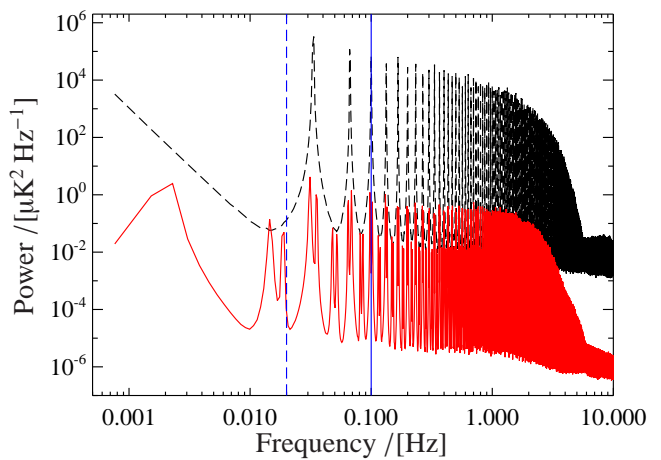}
\caption{Power vs frequency for temperature (in dashed black line) and polarization intensity (in solid red line)
in the case of half wave plate (HWP) steps of $\SI[per-mode=symbol]{11.25}{\degree\per\minute}$.
In our simulations, we focused on two noise knee frequencies: 0.02 and $\SI{0.1}{\hertz}$.}
\label{IPspectra}
\end{figure}
The payload will host two instruments: the Short Wavelength Instrument for the Polarization Explorer \citep[SWIPE;][]{2012SPIE.8452E..3FD}, which will map the sky in three frequency bands centered at 140, 220, and 
$\SI{240}{\giga\hertz}$; and the STRatospheric Italian Polarimeter \citep[STRIP,][]{2012SPIE.8446E..7CB}, which will survey the same sky region in two frequency bands centered at 43 and $\SI{90}{\giga\hertz}$. 

In this work we focus on the SWIPE bolometric polarimeter instrument, which is characterized by two symmetric orthogonally-placed focal planes hosting an overall number of 110 detectors per frequency. The peculiarity of SWIPE will be the presence of a half wave plate (HWP) polarization modulator as first optical element, followed by a $\SI{50}{\centi\meter}$ aperture refractive telescope, a beam-splitting polarizer, and finally the multimoded focal planes.
The SWIPE detectors are multimode spiderweb TES bolometers operating at $\SI{0.3}{\kelvin}$ \citep[see][]{2016JLTP..tmp....3G}.

In Fig.~\ref{LSPEmask} we show the sky region as seen by a subset of 18 LSPE-SWIPE detectors, which are arranged in three triples sparsely located in each of the two focal planes, and the corresponding angle coverage estimator, for one observation day. The maps are at HEALPix\footnote{http://healpix.sourceforge.net} resolution $N_{side}=128$ (27.5' per pixel, see \citealt{2005ApJ...622..759G}). In Fig.~\ref{IPspectra} we show the temperature and polarization intensity power as a function of frequency for one bolometer of the SWIPE instrument, assuming the nominal HWP steps of $\SI[per-mode=symbol]{11.25}{\degree\per\minute}$.

Because of the large number of detectors we expect that the noise cross-correlation may represent a critical issue and therefore, a proper treatment is  necessary. However, we highlight that the cross-correlation among the detectors must be accounted for in relation to the possible filtering of the data; the benefit of considering the noise common mode crucially depends on the magnitude of the low-frequency cut. For instance, we tested the extended ROMA code on the BOOMERanG real data set and we found no remarkable benefits with respect to the past analysis that completely neglected the noise cross-correlation. The low impact of the new treatment is due to the heavy filter of low-frequency data streams performed on the real data set, which has crudely cut out the information at large angular scales where the cross-correlated noise effect is more relevant. However, using simulated unfiltered data we found that a proper treatment of the noise cross-correlation results in considerably better maps and angular power spectrum estimates (the spectrum standard deviations have been reduced up to 20\%; see \citealt{AleThesis} and \citealt{2016JPhCS.689a2003B}).
In the following simulations, we first assume that no filter was applied on the data, hence we face a specific case where reasonable frequency cuts have been performed.

\subsection{Simulations}

We generate simulated TOD based on the SWIPE scanning strategy and polarimeter angles for one day of observations ($N_d = 8.64 \times 10^6$ samples per detector) and telescope elevation of 45 deg. In particular we choose an HWP stepping of $\SI[per-mode=symbol]{11.25}{\degree\per\minute}$, repeatedly scanning the range $\SI{0}{\degree}-\SI{78.75}{\degree}$, and we consider 18 detectors as described above.

We assume the cosmological parameters estimated by \Planck\ \citep{2015arXiv150201589P} with a tensor-to-scalar ratio $r=0.09$. The noise is simulated assuming the bolometer to be photon-noise limited with a white plateau of $\SI{15}{\micro\kelvin_{CMB}\sqrt{\second}}$ at higher frequencies and a $1/f^{\alpha}$ noise with $\alpha=2$ at lower frequencies,
as expected for the multimoded $\SI{140}{\giga\hertz}$ channel.
We focus on two knee frequencies, 0.02 and $\SI{0.1}{\hertz}$, the expected best and worst case.
For each knee frequency, we consider two different common-mode configurations:
\begin{enumerate}
  \item The cross-correlation is present in both the $1/f^{2}$ and white noise part of the spectrum. 
  In the former, the cross-correlated component is shared by all the detectors;
  in the latter we assume 10\% of cross-correlation among each triple and 1\% among any other detector,
  with respect to the auto noise spectrum.
  This situation is the analog to that found in the \boom\ analysis;\label{itm:uno}
\item The cross-correlation is limited to the $1/f^{2}$ part of the noise spectrum with no white noise common mode.\label{itm:due}
\end{enumerate}
\begin{figure}
\centering
\psfrag{l(l+1)C_lBB}[c][][3]{$\ell(\ell+1) C_{\ell}^{BB}/(2\pi)/[\si{\micro  \kelvin \squared}]$}
\psfrag{Multipolel}[c][][3]{Multipole $\ell$}
\psfrag{f_k=0.02Hz}[c][][3]{$f_k=\SI{0.02}{\hertz}$}
\includegraphics[angle=90,width=\columnwidth]{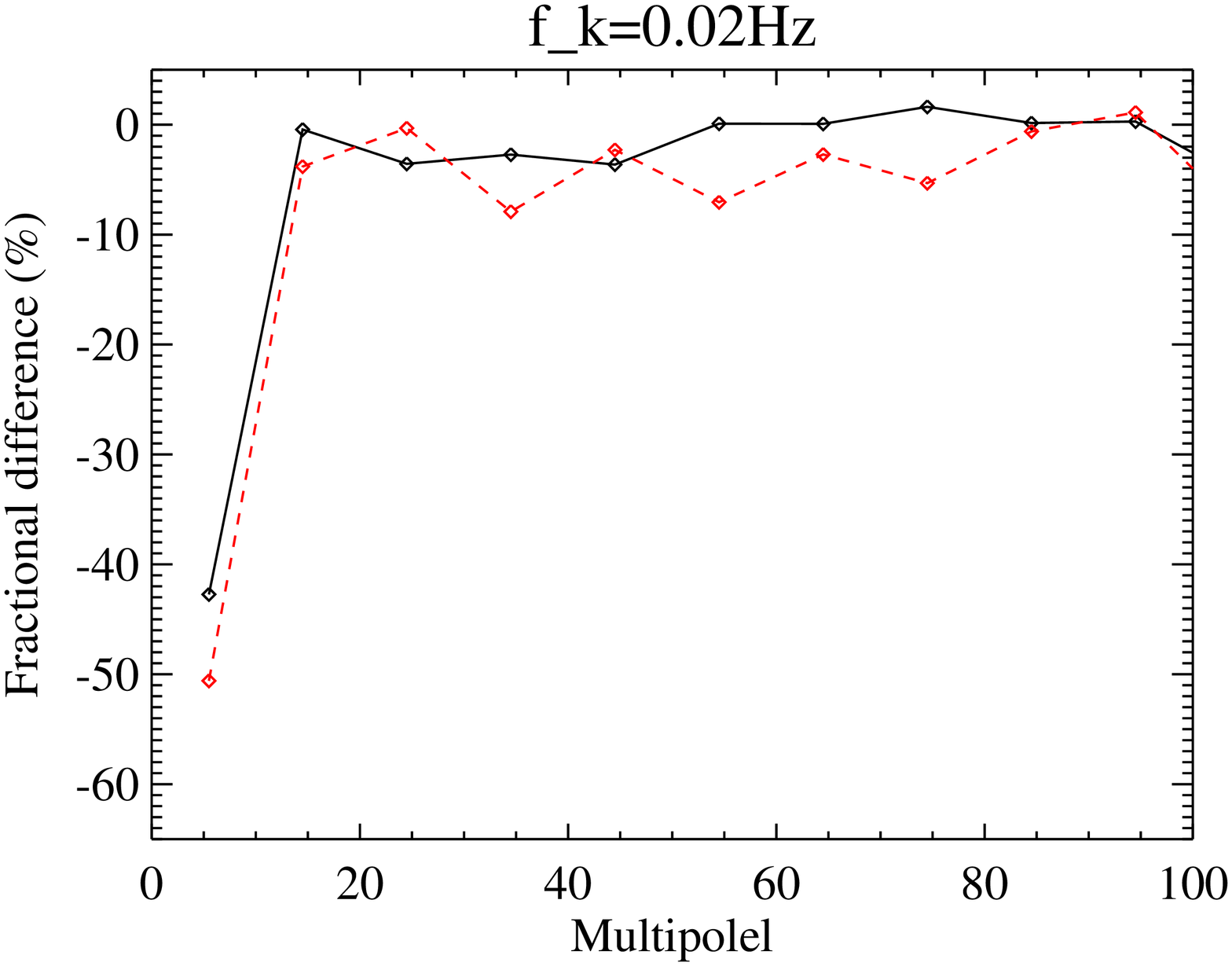}
\caption{
Fractional difference of the BB power spectra error bars from LSPE-SWIPE simulations estimated considering and neglecting the bolometer noise cross-correlation of case~\ref{itm:uno} (in dashed red line) and case~\ref{itm:due} (in solid black line) for a knee frequency $f_{k}=\SI{0.02}{\hertz}$.
}
\label{LSPE_0.02}
\end{figure}
\begin{figure}
\centering
\psfrag{l(l+1)C_lBB}[c][][3]{$\ell(\ell+1) C_{\ell}^{BB}/(2\pi) /[\si{\micro  \kelvin \squared}]$}
\psfrag{Multipolel}[c][][3]{Multipole $\ell$}
\psfrag{f_k=0.1Hz}[c][][3]{$f_k=\SI{0.1}{\hertz}$}
\includegraphics[angle=90,width=\columnwidth]{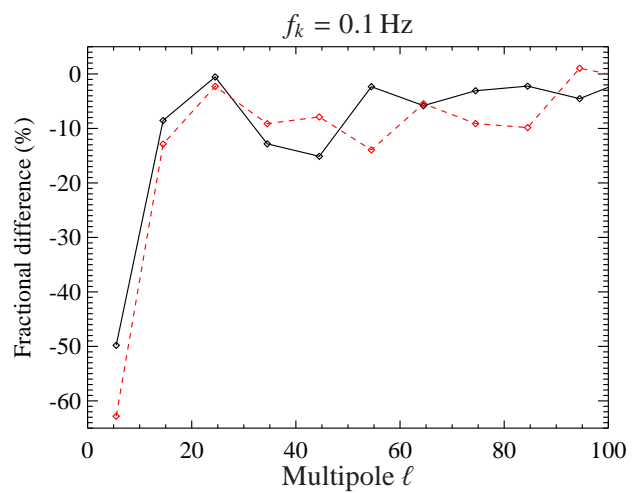}
\caption{
Fractional difference of the BB power spectra error bars from LSPE-SWIPE simulations estimated considering and neglecting the bolometer noise cross-correlation of case~\ref{itm:uno} (in dashed red line) and case~\ref{itm:due} (in solid black line) for a knee frequency $f_{k}=\SI{0.1}{\hertz}$.
}
\label{LSPE_0.1}
\end{figure}
In the four cases, we produced 50 signal-only, noise-only, and signal plus noise Monte Carlo simulated maps,
both taking into account or not the noise cross-correlation among detectors in the map-making code.
Then, we applied the MASTER power spectrum estimator \citep{2002ApJ...567....2H} to the maps.

It should be taken into account that, at large angular scales, the MASTER spectrum estimator method may not be the most convenient choice. As mentioned by \citet{2014MNRAS.440..957M}, at low multipoles a quadratic maximum likelihood estimator (QML) is preferable. 
Nonetheless, the aim of this work is to check the reliability  of an improved map-making code, which can also be successfully tested by using quicker but less accurate angular power spectra estimates.

In Fig.~\ref{LSPE_0.02} we show the fractional difference of BB power spectra error bars estimated considering and neglecting the noise cross-correlation in the algorithm, assuming the data set to be contaminated by the two noise configuration described above, for a knee frequency $f_{k}=\SI{0.02}{\hertz}$. The results for $f_{k}=\SI{0.1}{\hertz}$ are shown in Fig.~\ref{LSPE_0.1}.
 It is evident that neglecting the cross-correlation affects heavily the power spectra error bars.
For a knee frequency $f_{k}=\SI{0.02}{\hertz}$, the inclusion of the cross-correlation has the effect of reducing
the spectra error bars up to $40-50\%$. For $f_{k}=\SI{0.1}{\hertz}$, we find an improvement up to $50-60\%$. 

\subsection{Comparison with filtering techniques}

As mentioned above, to reduce the low-frequency noise contribution, it is a common choice to high pass the data stream at some cut frequency $f_c$. However, this method is not lossless, as  part of the signal information is filtered out as well.

While this option appears to be feasible when the low-frequency cutoff $f_c$ is $\SI{0.02}{\hertz}$ since the cosmological signal at lower frequencies is negligible, we found that
(see Fig.~\ref{IPspectra}) if $f_c=0.05$ or $\SI{0.1}{\hertz}$ the high-pass filter will cut the polarization intensity power of $\simeq 17$ and $25\%$ and the $I$ Stokes power of $\simeq 31$ and $48\%$, respectively.

To estimate the effect of data filtering on the power spectra error bars, we perform a similar analysis as above but with high-pass frequencies $f_c=0.02, 0.05$ and $\SI{0.1}{\hertz}$ and neglecting the cross-correlated noise in the analysis for the noise configuration of case~\ref{itm:due} and $f_k=\SI{0.1}{\hertz}$.

\begin{figure}
\centering
\psfrag{Multipolel}[c][][3]{Multipole $\ell$}
\includegraphics[angle=90,width=\columnwidth]{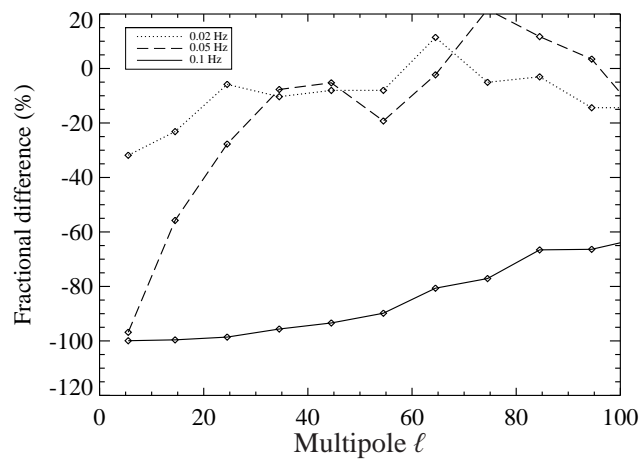}
\caption{
Fractional difference between the BB power spectrum error bars estimated including the noise cross-correlation and applying a low-frequency data filtering (neglecting the cross-correlated noise) at $f_c=\SI{0.02}{\hertz}$ in dotted line, $f_c=\SI{0.05}{\hertz}$ in dashed line and $f_c=\SI{0.1}{\hertz}$ in solid line.
The BB spectra correspond to the noise configuration of case~\ref{itm:due} with $f_k=\SI{0.1}{\hertz}$.
}
\label{filter}
\end{figure}

We compare these results with the power spectra error bars estimated accounting for cross-correlated noise with no data filtering. In Fig.~\ref{filter} we show the fractional difference of the error bars between the cases with cross-correlation and low-frequency filtering. It can be noticed that, by including the noise cross-correlation in the map-making code, we are recovering smaller BB power spectrum error bars compared to the case with data filtering. At very low multipoles, the improvement is up to 30\% and 100\% for the $f_c=\SI{0.02}{\hertz}$ and $f_c \geq \SI{0.05}{\hertz}$ cases, respectively.

It is not the aim of this work to forecast a suitable filtering strategy for LSPE-SWIPE in relation to the inclusion of the noise cross-correlation in the analysis. This would require a more accurate specifications of many parameters, such as telescope elevation, azimuth scan velocity and HWP velocity.
However, we stress that the improvements provided by the extended ROMA code may result in a less dramatic filtering of low-frequency data, thus preserving most of the cosmological information.

%%%%%%%%%%%%%%%%%%%%%%%%%%%%%%%%%%%%%%%%%%%%%%%%%%%%%%%%%%%%%%%%%%%%%%%%%%%%%%%%%%%%%%%%%%%%%%%%%%%%%%%%%%%%%%%%%%%
\section{Conclusions}\label{Conclusions}
We presented a new version of the ROMA map-making code extended to a possible cross-correlated noise component among the detectors of a CMB experiment.

This effect must be properly taken into account in the data analysis of any experiment aimed at detection of B-mode polarization, either from space, balloons or ground.

Among the next generation of CMB experiments, we focus on the forthcoming LSPE balloon-borne experiment, devoted to the accurate observation of CMB polarization at large angular scales with the primary aim to constrain the primordial B-mode polarization.

We applied the extended ROMA algorithm to simulated data of the LSPE-SWIPE 
instrument and found that the inclusion of cross-correlation provides spectra error bars smaller up to $50-60\%$, depending on the chosen knee frequency. 
We point out that this improvement could be crucial in constraining the $B$-mode polarization.

A very relevant issue is to compare these results with a possible low-frequency data filtering, commonly used for ground-based and balloon-borne experiments.
We found that accounting for cross-correlated noise
with no filtering is a viable option, which is certainly less crude than a cut of 
low-frequency streams, with the additional potential advantage of not
removing cosmological information at the largest scales.
%
%
%%%%%%%%%%%%%%%%%%%%%%%%%%%%%%%%%%%%%%%%%%%%%%%%%%%%%%%%%%%%%%%%%%%%%%%%%%%

\begin{acknowledgements} 
    We acknowledge the use of the HEALPix  package \citep{2005ApJ...622..759G}
    and of the FFTW library \citep{FFTW05}. We wish to thank Silvia Masi and Marina Migliaccio for useful suggestions and discussions, and LSPE collaboration for providing us with the LSPE-SWIPE pointing informations.
\end{acknowledgements}

\bibliographystyle{aa} % style aa.bst
\bibliography{roma} % your references roma.bib

\begin{thebibliography}{19}
\expandafter\ifx\csname natexlab\endcsname\relax\def\natexlab#1{#1}\fi

\bibitem[{{Bersanelli} {et~al.}(2012){Bersanelli}, {Mennella}, {Morgante},
  {Zannoni}, {Addamo}, {Baschirotto}, {Battaglia}, {Ba{\'o}}, {Cappellini},
  {Cavaliere}, {Cuttaia}, {Del Torto}, {Donzelli}, {Farooqui}, {Frailis},
  {Franceschet}, {Franceschi}, {Gaier}, {Galeotta}, {Gervasi}, {Gregorio},
  {Kangaslahti}, {Krachmalnicoff}, {Lawrence}, {Maggio}, {Mainini}, {Maino},
  {Mandolesi}, {Paroli}, {Passerini}, {Peverini}, {Poli}, {Ricciardi},
  {Rossetti}, {Sandri}, {Seiffert}, {Stringhetti}, {Tartari}, {Tascone},
  {Tavagnacco}, {Terenzi}, {Tomasi}, {Tommasi}, {Villa}, {Virone}, \&
  {Zacchei}}]{2012SPIE.8446E..7CB}
{Bersanelli}, M., {Mennella}, A., {Morgante}, G., {et~al.} 2012, in Society of
  Photo-Optical Instrumentation Engineers (SPIE) Conference Series, Vol. 8446,
  Society of Photo-Optical Instrumentation Engineers (SPIE) Conference Series,
  7

\bibitem[{{BICEP2/Keck and Planck Collaborations} {et~al.}(2015){BICEP2/Keck
  and Planck Collaborations}, {Ade}, {Aghanim}, {Ahmed}, {Aikin}, {Alexander},
  {Arnaud}, {Aumont}, {Baccigalupi}, {Banday}, \& et~al.}]{2015PhRvL.114j1301B}
{BICEP2/Keck and Planck Collaborations}, {Ade}, P.~A.~R., {Aghanim}, N.,
  {et~al.} 2015, Physical Review Letters, 114, 101301

\bibitem[{Buzzelli(2015)}]{AleThesis}
Buzzelli, A. 2015, Master's thesis, Universit{\`a} di Roma ``Tor Vergata'',
  Italy

\bibitem[{{Buzzelli} {et~al.}(2016){Buzzelli}, {Cabella}, {de Gasperis}, \&
  {Vittorio}}]{2016JPhCS.689a2003B}
{Buzzelli}, A., {Cabella}, P., {de Gasperis}, G., \& {Vittorio}, N. 2016,
  Journal of Physics Conference Series, 689, 012003

\bibitem[{{Cantalupo} {et~al.}(2010){Cantalupo}, {Borrill}, {Jaffe}, {Kisner},
  \& {Stompor}}]{2010ApJS..187..212C}
{Cantalupo}, C.~M., {Borrill}, J.~D., {Jaffe}, A.~H., {Kisner}, T.~S., \&
  {Stompor}, R. 2010, \apjs, 187, 212

\bibitem[{{de Bernardis} {et~al.}(2012){de Bernardis}, {Aiola}, {Amico},
  {Battistelli}, {Coppolecchia}, {Cruciani}, {D'Addabbo}, {D'Alessandro}, {De
  Gregori}, {De Petris}, {Goldie}, {Gualtieri}, {Haynes}, {Lamagna}, {Maffei},
  {Masi}, {Nati}, {Ng}, {Pagano}, {Piacentini}, {Piccirillo}, {Pisano},
  {Romeo}, {Salatino}, {Schillaci}, {Tommasi}, \&
  {Withington}}]{2012SPIE.8452E..3FD}
{de Bernardis}, P., {Aiola}, S., {Amico}, G., {et~al.} 2012, in Society of
  Photo-Optical Instrumentation Engineers (SPIE) Conference Series, Vol. 8452,
  Society of Photo-Optical Instrumentation Engineers (SPIE) Conference Series,
  3

\bibitem[{{de Gasperis} {et~al.}(2005){de Gasperis}, {Balbi}, {Cabella},
  {Natoli}, \& {Vittorio}}]{2005A&A...436.1159D}
{de Gasperis}, G., {Balbi}, A., {Cabella}, P., {Natoli}, P., \& {Vittorio}, N.
  2005, \aap, 436, 1159

\bibitem[{Frigo \& Johnson(2005)}]{FFTW05}
Frigo, M. \& Johnson, S.~G. 2005, Proceedings of the IEEE, 93, 216, special
  issue on ``Program Generation, Optimization, and Platform Adaptation''

\bibitem[{{G{\'o}rski} {et~al.}(2005){G{\'o}rski}, {Hivon}, {Banday},
  {Wandelt}, {Hansen}, {Reinecke}, \& {Bartelmann}}]{2005ApJ...622..759G}
{G{\'o}rski}, K.~M., {Hivon}, E., {Banday}, A.~J., {et~al.} 2005, \apj, 622,
  759

\bibitem[{{Gualtieri} {et~al.}(2016){Gualtieri}, {Battistelli}, {Cruciani}, {de
  Bernardis}, {Biasotti}, {Corsini}, {Gatti}, {Lamagna}, \&
  {Masi}}]{2016JLTP..tmp....3G}
{Gualtieri}, R., {Battistelli}, E.~S., {Cruciani}, A., {et~al.} 2016, Journal
  of Low Temperature Physics [\eprint[arXiv]{1602.07744}]

\bibitem[{{Hivon} {et~al.}(2002){Hivon}, {G{\'o}rski}, {Netterfield}, {Crill},
  {Prunet}, \& {Hansen}}]{2002ApJ...567....2H}
{Hivon}, E., {G{\'o}rski}, K.~M., {Netterfield}, C.~B., {et~al.} 2002, \apj,
  567, 2

\bibitem[{{Lyth} \& {Riotto}(1999)}]{1999PhR...314....1L}
{Lyth}, D.~H.~D.~H. \& {Riotto}, A.~A. 1999, \physrep, 314, 1

\bibitem[{{Masi} {et~al.}(2006){Masi}, {Ade}, {Bock}, {Bond}, {Borrill},
  {Boscaleri}, {Cabella}, {Contaldi}, {Crill}, {de Bernardis}, {de Gasperis},
  {de Oliveira-Costa}, {de Troia}, {di Stefano}, {Ehlers}, {Hivon}, {Hristov},
  {Iacoangeli}, {Jaffe}, {Jones}, {Kisner}, {Lange}, {MacTavish}, {Marini
  Bettolo}, {Mason}, {Mauskopf}, {Montroy}, {Nati}, {Nati}, {Natoli},
  {Netterfield}, {Pascale}, {Piacentini}, {Pogosyan}, {Polenta}, {Prunet},
  {Ricciardi}, {Romeo}, {Ruhl}, {Santini}, {Tegmark}, {Torbet}, {Veneziani}, \&
  {Vittorio}}]{2006A&A...458..687M}
{Masi}, S., {Ade}, P.~A.~R., {Bock}, J.~J., {et~al.} 2006, \aap, 458, 687

\bibitem[{{Molinari} {et~al.}(2014){Molinari}, {Gruppuso}, {Polenta},
  {Burigana}, {De Rosa}, {Natoli}, {Finelli}, \& {Paci}}]{2014MNRAS.440..957M}
{Molinari}, D., {Gruppuso}, A., {Polenta}, G., {et~al.} 2014, \mnras, 440, 957

\bibitem[{{Patanchon} {et~al.}(2008){Patanchon}, {Ade}, {Bock}, {Chapin},
  {Devlin}, {Dicker}, {Griffin}, {Gundersen}, {Halpern}, {Hargrave}, {Hughes},
  {Klein}, {Marsden}, {Martin}, {Mauskopf}, {Netterfield}, {Olmi}, {Pascale},
  {Rex}, {Scott}, {Semisch}, {Truch}, {Tucker}, {Tucker}, {Viero}, \&
  {Wiebe}}]{2008ApJ...681..708P}
{Patanchon}, G., {Ade}, P.~A.~R., {Bock}, J.~J., {et~al.} 2008, \apj, 681, 708

\bibitem[{{Planck Collaboration} {et~al.}(2015{\natexlab{a}}){Planck
  Collaboration}, {Ade}, {Aghanim}, {Arnaud}, {Ashdown}, {Aumont},
  {Baccigalupi}, {Banday}, {Barreiro}, {Bartlett}, \&
  et~al.}]{2015arXiv150201589P}
{Planck Collaboration}, {Ade}, P.~A.~R., {Aghanim}, N., {et~al.}
  2015{\natexlab{a}}, ArXiv e-prints [\eprint[arXiv]{1502.01589}]

\bibitem[{{Planck Collaboration} {et~al.}(2015{\natexlab{b}}){Planck
  Collaboration}, {Aghanim}, {Arnaud}, {Ashdown}, {Aumont}, {Baccigalupi},
  {Banday}, {Barreiro}, {Bartlett}, {Bartolo}, \& et~al.}]{2015arXiv150702704P}
{Planck Collaboration}, {Aghanim}, N., {Arnaud}, M., {et~al.}
  2015{\natexlab{b}}, ArXiv e-prints [\eprint[arXiv]{1507.02704}]

\bibitem[{{Tristram} {et~al.}(2011){Tristram}, {Filliard}, {Perdereau},
  {Plaszczynski}, {Stompor}, \& {Touze}}]{2011A&A...534A..88T}
{Tristram}, M., {Filliard}, C., {Perdereau}, O., {et~al.} 2011, \aap, 534, A88

\bibitem[{{Wright}(1996)}]{1996astro.ph.12006W}
{Wright}, E.~L. 1996, ArXiv Astrophysics e-prints [\eprint{astro-ph/9612006}]

\end{thebibliography}
\end{document}